\documentclass[a4paper,11pt]{article}
\pdfoutput=1

\usepackage{hyperref}
\usepackage{bm}

\usepackage{braket,slashed}
\usepackage{amsmath}
\usepackage{amssymb}

\usepackage[superscript]{cite}

\providecommand{\keywords}[1]
{
  \small	
  \textbf{\textit{Keywords---}} #1
}

\begin{document}

\title{Higgs-Induced Screening Mechanisms in Scalar-Tensor Theories}

\author{Clare Burrage$^1$, Peter Millington$^2$} 
\date{$^1$ School of Physics and Astronomy, University of Nottingham, Nottingham, NG7 2RD, UK\\
$^2$ Department of Physics and Astronomy, University of Manchester, Manchester, M13 9PL, UK}

\maketitle

%%%%%%%%%%%%%%%%%%%%

\begin{abstract}
We consider the theory of a light conformally coupled scalar field, i.e., one that is coupled directly to the Ricci scalar of the gravitational sector. This theory can be written equivalently as one of a light scalar that is coupled to the Standard Model of particle physics with a particular combination of Higgs-portal couplings. When the conformal coupling function contains terms that are linear and quadratic in the conformally coupled scalar, we find that the effective mass of the light propagating mode and its coupling to matter fields, obtained after expanding around a minimum of the classical potential, depend on the energy density of the background environment.  This is despite the absence of non-linear terms  in the original equation of motion for the light conformally coupled field. Instead, we find that the non-linearities of the prototype Higgs potential are communicated to the light mode. In this way, we present a novel realisation of screening mechanisms, in which light degrees of freedom coupled to the Standard Model are able to avoid experimental constraints through environmental and thin-shell effects. 
\end{abstract}

%%%%%%%%%%%%%%%%%%%%


\newpage

%%%%%%%%%%%%%%%%%%%%

\section{Introduction}

Light scalar fields are a popular candidate for physics beyond the Standard Model (SM), with significant motivation coming from theories of dark matter \cite{Turner:1983he, Press:1989id, Sin:1992bg, Goodman:2000tg, Peebles:2000yy, Hu:2000ke, Arbey:2001qi, Amendola:2005ad, Schive:2014dra, Hui:2016ltb, Marsh:2015wka, Ferreira:2020fam}, dark energy \cite{Copeland:2006wr, Joyce:2014kja, Bull:2015stt} and modified gravity \cite{Clifton:2011jh, Heisenberg:2018vsk, Kobayashi:2019hrl, CANTATA:2021ktz}. 
However, there is, as yet, no evidence of new, light scalars coupled to the SM particles.

One way to explain the lack of evidence of new scalars is to tune the coupling of the scalar to the SM to be small~\cite{Adelberger:2003zx}.  If we wish to avoid this tuning, there are currently two options available. The first is to couple the scalar field conformally to a fully scale-invariant SM Lagrangian. In this case, a symmetry suppresses all interactions between the scalar field and the fermions of the SM \cite{Wetterich:1987fm, Buchmuller:1988cj, Shaposhnikov:2008xb, Shaposhnikov:2008xi, Blas:2011ac, Garcia-Bellido:2011kqb, Bezrukov:2012hx, Henz:2013oxa, Rubio:2014wta, Brax:2014baa, Karananas:2016grc, Ferreira:2016kxi, Ferreira:2016vsc}. However, to preserve scale invariance, the theory requires an unusual approach to renormalisation \cite{Coleman:1973jx, Englert:1976ep, Ghilencea:2015mza, Ghilencea:2016ckm, Ferreira:2018itt}. A second option is offered by theories with environmentally dependent screening, where observable effects, such as fifth forces, can be naturally suppressed in the neighbourhood of experiments \cite{Joyce:2014kja, Koyama:2015vza, Burrage:2017qrf, Ishak:2018his}.  The cost paid for this behaviour is that the equations of motion of the theory must be non-linear. These non-linearities can involve non-trivial self-interactions of the scalar, non-linear matter couplings or non-canonical kinetic terms, or a combination of all three. 

Renormalisable self-interactions are not forbidden for scalar field theories. Indeed, the one scalar field that we have observed --- the Higgs field --- is thought to possess non-trivial quartic self-interactions, which, along with the quadratic term of the Higgs potential, are vital for electroweak symmetry breaking. Theories of non-linear light scalar fields with environmentally dependent behaviour are often referred to as screened scalars. The commonly studied models with screening, including chameleon \cite{Khoury:2003aq, Khoury:2003rn}, symmetron \cite{Hinterbichler:2010es, Hinterbichler:2011ca} and Vainshtein screening \cite{Vainshtein:1972sx, Dvali:2000hr, Nicolis:2008in}, have all, to varying degrees, faced challenges about their naturalness, and whether the light masses can be protected from corrections due to interactions with heavier fields.  

In this work, we attempt to address these challenges to screened theories by considering a scalar field with a small mass, which couples to the SM conformally, i.e., via a non-minimal coupling to the Ricci scalar.  This means that SM particles move on geodesics of a metric that is conformally re-scaled by a function of the additional scalar field.  Such couplings naturally arise in UV theories with extra dimensions, e.g., string-theory dilatons\cite{Damour:1994zq, Damour:2002nv, Gasperini:2001pc, Damour:2002mi}, which may be screened,\cite{Brax:2008hh, Brax:2010gi} as well as theories of modified gravity, such as $f(R)$ theories.\cite{Sotiriou:2008rp} In previous work, and by virtue of the scale-symmetry breaking provided by the quadratic term in the Higgs potential, we have shown how models involving conformally coupled scalars can be rewritten as Higgs-portal models \cite{Burrage:2018dvt}, being related by the Weyl rescaling of the metric from the so-called Jordan frame to the Einstein frame.  This is to say that there is a field basis in the Einstein frame in which the scalar only interacts directly with the Higgs (at dimension four) and has no direct couplings to the fermions of the SM.  Fifth-force couplings of the light degree of freedom to the SM fermions can then be seen to arise as a result of mixing with the Higgs, or after diagonalising this mixing~\cite{Burrage:2018dvt}.  An equivalent result can be obtained directly in the Jordan frame, wherein the fifth-force coupling to SM fermions arises after diagonalising the kinetic mixing of the conformally coupled scalar and the graviton~\cite{Copeland:2021qby}.

The Higgs portal offers the lowest-dimension, renormalisable portal by which to couple new fields (also known as hidden sectors) to the SM \cite{Binoth:1996au, Patt:2006fw, Schabinger:2005ei, Englert:2011yb}. Light scalars coupled through the Higgs portal have received much recent attention \cite{Bauer:2020nld, Beacham:2019nyx}, but the possibilities of screening through non-linearities, which are naturally present, have largely been  overlooked, with a few exceptions (see, e.g., Ref.~\citen{Brax:2021rwk}).

In the following section, we introduce our model, both in terms of its conformal and Higgs-portal couplings.   In Section \ref{sec:intout},  we  see the first signs of environmental dependence through the expectation value of the conformally coupled scalar field, which will be seen to depend on the local energy density. We derive the effective equation of motion for the light scalar mode by expanding to leading order in fluctuations around a density-dependent minimum of the classical potential. In Section \ref{sec:screenign}, we then show how this environmental dependence leads to suppression of the interactions between the light mode and matter, and how this leads to dynamical screening of the fifth forces sourced by massive compact objects. We discuss the implications and limitations  of these results further in Section \ref{sec:dis}, and conclude in Section \ref{sec:conc}.

%%%%%%%%%%%%%%%%%%%%

\section{Conformal couplings and the Higgs portal}

In our previous work, Ref.~\citen{Burrage:2018dvt} (see also Refs.~\citen{Copeland:2021qby, SevillanoMunoz:2022tfb}), we showed that conformally coupled theories were equivalent, at tree level, to Higgs-portal models. 
We started with a generic action for a conformally coupled scalar-tensor theory, written in the Einstein frame as
\begin{equation}
\label{eq:startingaction}
S\ = \ \int\!{\rm d}^4x\,\sqrt{-\,\tilde{g}}\bigg[\frac{M_{\rm Pl}^2}{2}\,\tilde{\mathcal{R}}\:-\:\frac{1}{2}\,\tilde{g}^{\mu\nu}\,\partial_{\mu}\chi\,\partial_{\nu}\chi\:-\:V(\chi)\bigg]\:+\:S_{\rm SM}\big[A^2(\chi)\tilde{g}_{\mu\nu},\{\psi\}\big]\;,
\end{equation}
where the light scalar $\chi$  has a canonical kinetic term and a potential $V(\chi)$. $\tilde{\mathcal{R}}$ denotes the Ricci scalar for the Einstein frame metric $\tilde{g}_{\mu\nu}$, and $M_{\rm Pl} $ is the Planck mass. The term $S_{\rm SM}$ is the SM action, whose fields are indicated by $\{\psi\}$.  These fields move on geodesics of the Jordan-frame metric $g_{\mu\nu}=A^2(\chi)\tilde{g}_{\mu\nu}$. We work throughout with signature convention $(-,+,+,+)$.

We write a toy SM (with one fermion $\psi$ and a real prototype of the Higgs field $\phi$) in terms of the Jordan-frame metric as
\begin{align}
S_{\rm SM}[g_{\mu\nu},\{\psi\}]\ &=\ \int\!{\rm d}^4x\,\sqrt{-\,g}\bigg[-\:\frac{1}{2}\,g^{\mu\nu}\,\partial_{\mu}\phi\,\partial_{\nu}\phi\:+\:\frac{1}{2}\,\mu^2\,\phi^2\:-\:\frac{\lambda}{4!}\,\phi^4\:\nonumber\\&\qquad -\:\frac{3}{2}\,\frac{\mu^4}{\lambda} -\:\bar{\psi}ie_{a}^{\mu}\gamma^{a}\overset{\leftrightarrow}{\partial}_{\mu}\psi\:-\:y\,\bar{\psi}\phi\psi\bigg]\;,
\label{eq:sm}
\end{align}
We include the  constant term $-3\mu^4/2\lambda$ to set the energy of the ``Higgs'' potential at its minima to zero.  This sets to zero any contribution to the cosmological constant that could arise from the Higgs field after the would-be electroweak symmetry breaking. 

The scalar coupling to matter can be made explicit  by rewriting the theory in terms of the  Einstein-frame metric $\tilde{g}_{\mu\nu}$. The action is then
\begin{align}
S_{\rm SM}[A^2(\chi)\tilde{g}_{\mu\nu},\{\psi\}]\ &=\ \int\!{\rm d}^4x\,\sqrt{-\,\tilde{g}}\bigg[-\:\frac{1}{2}\,A^2(\chi)\tilde{g}^{\mu\nu}\,\partial_{\mu}\phi\,\partial_{\nu}\phi\:\nonumber\\
&\quad +\:\frac{1}{2}\,A^4(\chi)\,\mu^2\,\phi^2-\:\frac{\lambda}{4!}\,A^4(\chi)\,\phi^4\:-\:\frac{3}{2}\,A^4(\chi)\,\frac{\mu^4}{\lambda}\:\nonumber\\&\quad -\:A^2(\chi)\,\bar{\psi}i\overset{\leftrightarrow}{\slashed{\partial}}\psi\:-\:y\,A^4(\chi)\,\bar{\psi}\phi\psi\bigg]\;.
\end{align}
We note here the explicit appearance of the coupling function $A(\chi)$.

After redefining the Higgs and fermion fields according to their classical scaling dimensions as
\begin{equation}
\label{eq:phiredef}
\tilde{\phi}\ \equiv\ A(\chi)\phi\;,\qquad \tilde{\psi}\ \equiv\ A^{3/2}(\chi)\psi\;,
\end{equation}
our toy SM Lagrangian becomes
\begin{align}
\label{eq:LSMEframe}
\tilde{\mathcal{L}}\ &=\ -\:\frac{1}{2}\,\tilde{g}^{\mu\nu}\,\partial_{\mu}\tilde{\phi}\,\partial_{\nu}\tilde{\phi}\:+\:\tilde{g}^{\mu\nu}\,\tilde{\phi}\,\partial_{\mu}\tilde{\phi}\,\partial_{\nu}\ln A(\chi)\:\nonumber\\
&\qquad -\:\frac{1}{2}\,\tilde{g}^{\mu\nu}\,\tilde{\phi}^2\,\partial_{\mu}\ln A(\chi)\,\partial_{\nu}\ln A(\chi)\nonumber\\
&\qquad+\:\frac{1}{2}\,\mu^2\,A^2(\chi)\,\tilde{\phi}^2\:-\:\frac{\lambda}{4!}\,\tilde{\phi}^4\:-\:\frac{3}{2}\,A^4(\chi)\,\frac{\mu^4}{\lambda}\:\nonumber\\
&\qquad -\:\bar{\tilde{\psi}}i\overset{\leftrightarrow}{\tilde{\slashed{\partial}}}\tilde{\psi}\:-\:y\,\bar{\tilde{\psi}}\tilde{\phi}\tilde{\psi}\;,
\end{align}
where $\tilde{\slashed{\partial}}\equiv \tilde{e}_{a}^{\mu}\gamma^a\partial_{\mu} = A^{-1}(\chi)e_{a}^{\mu}\gamma^a\partial_{\mu}$ and the antisymmetrisation of the kinetic term with $\overset{\leftrightarrow}{{\partial}_{\mu}}=\tfrac{1}{2}(\overset{\rightarrow}{{\partial}_{\mu}}-\overset{\leftarrow}{{\partial}_{\mu}})$ avoids the appearance of the spin connection in the Lagrangian (see, e.g., Ref.~\citen{Ferreira:2016kxi}).

From equation (\ref{eq:LSMEframe}), we see that the light  scalar does not couple directly to fermions, and instead only couples to the Higgs through the `Higgs portal', a coupling which depends on the Higgs mass $\mu$.  This is unsurprising as the Higgs mass  is the only explicit mass scale in our toy  SM. In previous work, we showed how this coupling leads to tree-level fifth forces due to the mixing between the light scalar and the Higgs, and how this fifth force can be suppressed if part, or all, of the Higgs mass scale arises dynamically \cite{Burrage:2018dvt}. A similar result can be obtained directly in the Jordan frame~\cite{Copeland:2021qby}.

In order to study this theory further, we write the coupling function as a power series of the form
\begin{equation}
\label{eq:A}
A^2(\chi)\ =\ 1\:+\:b\,\frac{\chi}{M}\:+\:c\,\frac{\chi^2}{M^2}\:+\:\mathcal{O}\bigg(\frac{\chi^3}{M^3}\bigg)\;,
\end{equation}
where  $b$ and $c$ are dimensionless constants, and $M$ is a mass scale. The latter controls the strength of the interaction between the scalar, the $\chi$ field and matter, and could be considered as the cut-off of the theory. Equation~(\ref{eq:A}) can be considered as the leading-order approximation to the true form of the coupling function. For example, in dilaton models, the coupling function would be a series of powers of exponential functions. However, as long as $\chi \ll M$, our calculations will remain valid. We also include a  mass term in the potential for the $\chi$ field, taking
\begin{equation}
\label{eq:chipot}
V(\chi)\ =\ \frac{1}{2}\,\mu_{\chi}^2\,\chi^2\;.
\end{equation}
More complicated potentials are, of course, allowed and may lead to a more varied phenomenology. However, we will see that even with this minimal choice, which might naively be expected to lead to linear equations of motion for the conformally coupled scalar, the interactions that the conformally coupled scalar obtains with the Higgs field will lead to non-linearities that are sufficient to induce screening mechanisms for the fifth-force mediating light degree of freedom.

Defining
\begin{equation}
\label{eq:chiredef}
\tilde{\chi}\ \equiv\ \bigg(1\:+\:\frac{b^2\tilde{\phi}^2}{4M^2}\bigg)^{1/2}\chi\;,
\end{equation}
to approach canonical normalisation for the $\chi$ field, we have (keeping terms up to order $\tilde{\chi}^2/M^2$ and $\tilde{\phi^2}/M^2$)
\begin{align}
\label{eq:EFTL}
\tilde{\mathcal{L}}\ &=\ -\:\frac{1}{2}\,\tilde{g}^{\mu\nu}\,\partial_{\mu}\tilde{\chi}\,\partial_{\nu}\tilde{\chi}\:-\:\frac{1}{2}\,\tilde{g}^{\mu\nu}\,\partial_{\mu}\tilde{\phi}\,\partial_{\nu}\tilde{\phi}\:\nonumber\\
&\qquad +\:\frac{1}{2}\,\tilde{g}^{\mu\nu}\,\frac{\tilde{\phi}}{M}\left(b+2c\,\frac{\tilde{\chi}}{M}-b^2\,\frac{\tilde{\chi}}{2 M}\right)\,\partial_{\mu}\tilde{\phi}\,\partial_{\nu}\tilde{\chi}\nonumber\\
&\qquad+\:\frac{1}{2}\,\mu^2\,\tilde{\phi}^2\left(
1+b\,\frac{\tilde{\chi}}{M}+c\,\frac{\tilde{\chi}^2}{M^2}\right)\:-\:\frac{\lambda}{4!}\,\tilde{\phi}^4\:\nonumber\\
&\qquad -\:\frac{3}{2}\,\frac{\mu^4}{\lambda}\left(1+2b\,\frac{\tilde{\chi}}{M}+2c\,\frac{\tilde{\chi}^2}{M^2}+b^2\,\frac{\tilde{\chi}^2}{M^2}\right)\nonumber\\
&\qquad-\:\frac{1}{2}\mu_{\chi}^2\tilde{\chi}^2\left(1-\frac{b^2\tilde{\phi}^2}{4 M^2}\right)\:\bar{\tilde{\psi}}i\overset{\leftrightarrow}{\tilde{\slashed{\partial}}}\tilde{\psi}\:-\:y\,\bar{\tilde{\psi}}\tilde{\phi}\tilde{\psi}\:+\:\cdots\;.
\end{align}
This is a Higgs-portal model, where the portal couplings, of the form $(\alpha_{hs} \tilde{\chi} +\lambda_{hs} \tilde{\chi}^2)\tilde{\phi}^2$, are\footnote{We use the subscripts $h$ and $s$ here, in line with the literature, to indicate the Higgs field and the scalar field that has been added to the SM.}
\begin{align}
    \alpha_{hs}&= \frac{b\mu^2}{2M}\;,\\
    \lambda_{hs} &= \frac{c \mu^2}{2M^2} +\frac{b^2 \mu_{\chi}^2}{8M^2}\;.
\end{align}
This relationship between the portal couplings may not appear to be an obvious choice at first sight, but we have seen how it arises due the the nature of the conformal coupling. We note that, in addition to the Higgs-portal couplings, there are also kinetic mixing terms between the Higgs and the light scalar.  These kinetic mixings give rise to fifth forces between matter fields that are suppressed compared to the mass mixings that arise from the portal couplings, given the low momentum exchanges involved.

The potential terms that only include  the light scalar are
\begin{equation}
    V(\tilde{\chi})= \frac{1}{2}\mu_{\chi}^2\tilde{\chi}^2+\frac{3}{2}\,\frac{\mu^4}{\lambda}\left(1+2b\,\frac{\tilde{\chi}}{M}+2c\,\frac{\tilde{\chi}^2}{M^2}+b^2\,\frac{\tilde{\chi}^2}{M^2}\right)\;,
\end{equation}
where the second term (in curved brackets) arises due to the conformal coupling to the term which sets the energy of the minima of the Higgs potential to zero in the would-be electroweak symmetry breaking vacuum, thus subtracting the cosmological constant contribution. Note that this also gives rise to non-linear terms in the potential for $\tilde{\chi}$, but these are suppressed for $\tilde{\chi}/M\ll 1$.

%%%%%%%%%%%%%%%%%%%%

\section{Equation of motion for the fifth-force mediator}
\label{sec:intout}

We now proceed to derive the effective equation of motion of the fifth-force mediator. We do so by performing a mean-field approximation, expanding in fluctuations around a minimum of the classical potential. This leads to a mass mixing between the Higgs and $\tilde{\chi}$ fields, which, when diagonalised, gives rise to a direct coupling of the light mode to the matter source (here, the single Dirac fermion). We will see, however, that the effective mass and matter coupling strengths of this light mode depend on the ambient matter density, as a result of the original mixing between the Higgs and $\chi$ fields. In this way, the non-linearities of the Higgs potential are communicated to the dynamics of the fifth-force mediator, leading to the screening effects that we describe in Section~\ref{sec:screenign}.

%%%%%%%%%%%%%%%%%%%%

\subsection{Minima of the potential}

The full Einstein-frame potential for the fields $\tilde{\chi}$, $\tilde{\phi}$ and $\tilde{\psi}$ in the Lagrangian of equation (\ref{eq:EFTL}) is
\begin{align}
  \tilde{V}(\tilde{\chi},\tilde{\phi},\tilde{\psi}) =& -\:\frac{1}{2}\,\mu^2\,\tilde{\phi}^2\left(
1+b\,\frac{\tilde{\chi}}{M}+c\,\frac{\tilde{\chi}^2}{M^2}\right)\:+\:\frac{\lambda}{4!}\,\tilde{\phi}^4\:\nonumber\\
&\qquad +\:\frac{3}{2}\,\frac{\mu^4}{\lambda}\left(1+2b\,\frac{\tilde{\chi}}{M}+2c\,\frac{\tilde{\chi}^2}{M^2}+b^2\,\frac{\tilde{\chi}^2}{M^2}\right)\nonumber\\
&\qquad+\:\frac{1}{2}\mu_{\chi}^2\tilde{\chi}^2\left(1-\frac{b^2\tilde{\phi}^2}{4 M^2}\right)\:
+\:y\,\bar{\tilde{\psi}}\tilde{\phi}\tilde{\psi}\;.
\label{eq:fullpot}
\end{align}
We now  assume that we are working in an environment with a background density of fermions. When these fermions are non-relativistic, their energy-momentum tensor can be related directly to the mass term in the Lagrangian, such that we can write $\rho_{\psi} = y \tilde{\phi}_m\langle \bar{\tilde{\psi}}\tilde{\psi}\rangle$, where $\tilde{\phi}_m$ is the value of $\tilde{\phi}$ at the minimum of the potential. This expression can be interpreted as a mean-field approximation for the non-relativistic limit of the fermion energy-momentum tensor, valid when taking the classical limit in the case of high occupation numbers.
After making this assumption for the behaviour of the fermions, we can study the behaviour of the scalar fields in this environment. Varying equation~(\ref{eq:fullpot}) with respect to $\tilde{\phi}$ and $\tilde{\chi}$, we find equations for the values $\tilde{\phi}_m$ and $\tilde{\chi}_m$ of the fields at the minima of the potential. These are
 \begin{equation}
     \frac{\mu^2\tilde{\phi}_m^4}{v^2}-\tilde{\phi}_m^2\left[\mu^2\left(1+\frac{b\tilde{\chi}_m}{M}+\frac{c\tilde{\chi}_m^2}{M^2}\right)+\frac{b\mu^2_{\chi}\tilde{\chi}_m^2}{4M^2}\right]+\rho_{\psi}=0\;,
 \end{equation}
 \begin{equation}
     \tilde{\chi}_m\left[\frac{\mu^2v^2}{M^2}\left(c-\frac{c\tilde{\phi}_m^2}{v^2}+\frac{b^2}{2}\right)+\mu^2_{\chi}\left(1-\frac{b^2\tilde{\phi}_m^2}{4M^2}\right)\right]=\frac{b\mu^2v^2}{2M}\left(1-\frac{\tilde{\phi}_m^2}{v^2}\right)\;,
 \end{equation}
where we have set $v^2 =6\mu^2/\lambda$. Keeping terms only to order $1/M^2$,   assuming that  $v \ll M$ and taking the mass scale of the light scalar to be much smaller than the mass scale of the Higgs, i.e., $\mu_{\chi}\ll \mu$,  we can solve these equations to find
\begin{align}
    \tilde{\phi}_m^2 &=v^2\left[1+\frac{b\tilde{\chi}_m}{M}+\frac{c\tilde{\chi}^2_m}{M^2} +\frac{b^2 \mu_{\chi}^2 \tilde{\chi}^2_m}{4\mu^2M^2}\right. \nonumber \\
    &\phantom{=}-\left.\frac{\rho_{\psi}}{\mu^2v^2}\left(1-\frac{b\tilde{\chi}_m}{M}-\frac{c\tilde{\chi}^2_m}{M^2}+\frac{b^2\tilde{\chi}^2_m}{M^2}-\frac{b^2 \mu_{\chi}^2 \tilde{\chi}^2_m}{4\mu^2M^2}\right)\right]\;,
    \label{eq:hmin}
\end{align}
\begin{equation}
    \frac{\tilde{\chi}_m}{M}= -\frac{b \rho_{\psi}}{2\mu_{\chi}^2M^2 +(2c-b^2)\rho_{\psi}}\;.
    \label{eq:chimin}
\end{equation}
We note that one might expect the terms  proportional to $\rho_{\psi}/(\mu^2v^2)$ inside the square bracket in equation (\ref{eq:hmin}) to be negligibly small.  In fact, it is important to keep them in order to determine the minimum for $\tilde{\chi}$ correctly (and  for computing the effective potential for the light mode in the next section), as leading-order terms cancel. The cancellation of the leading-order terms is a direct consequence of our choice to set the contribution of the Higgs field to the cosmological constant to zero --- notice that there is no contribution to $\tilde{\chi}_m/M$ proportional to $\mu^2v^2$, as one would otherwise expect.

In equation (\ref{eq:chimin}), we see the first signs of environmental dependence in this theory, as  the minimum for $\tilde{\chi}$ varies significantly depending on whether the environmental density is greater or smaller than a critical density $\rho_{\rm crit}=2\mu_{\chi}^2M^2/(2c-b^2)$. The limiting cases are 
\begin{equation}
    \frac{\tilde{\chi}_{m}}{M}= \left\{ \begin{array}{lc}
    -\frac{b\rho_{\psi}}{2M^2\mu_{\chi}^2}\;, & \mbox{ if } \rho_{\psi} \ll \rho_{\rm crit}\;,\\
    \frac{b}{b^2-2c}\;, & \mbox{ if } \rho_{\psi} \gg \rho_{\rm crit}\;.
    \end{array}\right.
    \label{eq:chiminlimit}
    \end{equation}
We note that a small tuning of our dimensionless constants $b$ and $c$ is required to ensure that $\tilde{\chi}_m < M$ and our theory remains  well defined in high-density environments. This means that it is not possible for the coupling function $A^2(\chi)$ to be a pure exponential, as if $b=1$ and $c=1/2$ in equation (\ref{eq:A}), then equation (\ref{eq:chiminlimit}) implies that $\tilde{\chi}_m$ diverges in high density environments.

We will explore the two regimes of behaviour that can be seen in equation (\ref{eq:chiminlimit}) further below. It is important to recognise, however, that this phenomenology is only possible if $\mu_{\chi}$ and therefore $\rho_{\rm crit}$ are non-zero.  For the field to remain truly massless requires a symmetry, e.g., scale or shift symmetry. In the absence of such a symmetry, a mass for the light scalar will be generated by quantum effects, and the calculations presented in this work will apply.

%%%%%%%%%%%%%%%%%%%%

\subsection{Equation of motion}

Many  experiments  that search for light scalars, e.g., fifth-force experiments,  are performed at energies well below that of the electroweak scale. At these low energies, we can expand around the minima of the classical potential and ignore fluctuations of heavy modes, with masses of order the electroweak scale.

We will do this by performing a mean-field expansion, under the assumption that the heavy field is slowly varying, and will consider the equation of motion for fluctuations of the light mode to first order. This is to say that we perform both a zeroth-order semi-classical approximation and a zeroth-order gradient expansion. The former amounts to neglecting corrections generated by integrating out the heavy fluctuations.\footnote{Integrating out heavy fluctuations around the classical (saddle-point) configurations will induce radiative corrections, and generate effective operators involving the light mode that carry additional electroweak-scale suppression. The latter are expected to be subdominant in the low-energy limit relevant to fifth forces. One could proceed to compute the evolution of the reduced density matrix for the light mode using the Feynman-Vernon influence functional to derive the relevant master equation. Integrating out the fluctuations in the heavy mode would then lead to non-local effective operators. In  Ref.~\citen{Burrage:2018pyg}, we performed this calculation for a closely related model, finding that corrections to the field evolution beyond the semi-classical contributions can be associated with the expected loop diagrams. While radiative corrections require fine tuning, loop corrections that depend on the spatial variation of the classical background field cannot be eliminated. However, these quantum corrections are not expected to be large when the semi-classical mean field is slowly varying compared to the Compton wavelength of the field\cite{Burrage:2021nys}, as we assume in this work.}
The latter amounts to neglecting gradients in the mean fields and terms with higher-order derivatives.

We perturb the $\tilde{\phi}$ and $\tilde{\chi}$ fields around the field values that minimise the potential, given in equations (\ref{eq:hmin}) and (\ref{eq:chimin}), writing $\tilde{\chi}=\tilde{\chi}_m +\delta\tilde{\chi}$ and $\tilde{\phi}=\tilde{\phi}_m +\delta\tilde{\phi}$. We keep terms in the equations of motion only to first order in perturbations. We find that mass terms in the equations of motion mix fluctuations of the two fields,  meaning that the heavy mode of the theory --- the ``Higgs'' boson --- does not directly correspond to fluctuations of the Higgs field $\delta\tilde{\phi}$. The mixing between the fields can be expressed in terms of a mixing angle $\theta$. Assuming that the mixing angle $\theta$ is small (large mixing angles are excluded by collider searches \cite{Bauer:2020nld}), we find that,  keeping terms only to order $1/M^2$, 
\begin{equation}
    \theta \approx-\frac{v}{2M}\left[b+\frac{(4c-b^2)\tilde{\chi}_m}{2M}+\frac{\rho_{\psi}}{\mu^2v^2}\left(b+\frac{2c\tilde{\chi}_m}{M}-\frac{5b^2\tilde{\chi}_m}{2M}\right)\right]\;.
\end{equation}
Herein, we have again neglected terms of order $\mu_{\chi}^2/\mu^2$, but we keep terms to first order in $\rho_{\psi}/(\mu^2v^2)$, since leading-order terms cancel in the calculation of the effective mass and coupling constant for the light mode.

We can now identify the heavy and light mass eigenstates in our theory, $h$ and $s$, respectively (working in this field basis removes non-derivative interactions between the fields in the equations of motion).  These are defined as 
\begin{align}
    h&= \delta\tilde{\phi}\cos\theta  + \delta\tilde{\chi}\sin \theta\;, \label{eq:h}\\
    s&=  \delta\tilde{\chi}\cos\theta - \delta\tilde{\phi}\sin \theta\;.\label{eq:s}
\end{align}
We will obtain an effective potential for the light mode $s$ by inverting equations (\ref{eq:h}) and (\ref{eq:s}) to write $\delta\tilde{\phi}$ and $\delta\tilde{\chi}$ in terms of $h$ and $s$, and substituting these expressions into the equations of motion for the fields. We then neglect derivatives of $h$, assuming that we are considering sufficiently low-energy experiments that the heavy mode is not perturbed from the minimum of the field potential. The resulting equation of motion for $s$ is
\begin{equation}    
\left(1+\frac{b^2v^2}{8M^2}\right)\Box  s = m_{\rm eff}^2 s+\frac{\beta(\rho_{\psi})}{M}\delta\rho_{\psi}\;.
\label{eq:scalareom}
\end{equation}
We note that there is no density dependence in the leading corrections to the kinetic terms. Hereafter, we omit terms that are suppressed by $v^2/M^2$, which could otherwise be re-scaled into the effective mass and coupling constants
\begin{equation}
    m_{\rm eff}^2 =\mu_{\chi}^2 +\frac{(4c-b^2)\rho_{\psi}}{4M^2}+ \mathcal{O}\left(\frac{v^2}{M^2}\right)\;,
    \label{eq:meff}
\end{equation}
and
\begin{equation}
    \beta(\rho_{\psi})= \frac{2b\mu_{\chi}^2 M^2}{2\mu_{\chi}^2M^2 +(2c-b^2)\rho_{\psi}}+\mathcal{O}\left(\frac{\rho_{\psi}}{\mu^2v^2}\right)\;.
\end{equation}

In low-density environments, the mass of the light scalar remains small in the leading semi-classical approximation.  This is a consequence of the choice to set the Higgs contribution to the cosmological constant to zero in the Jordan frame, resulting in the subtraction of $3\mu^4/(2\lambda)$ from the Higgs potential in equation (\ref{eq:sm}).\footnote{This article does not offer a solution to the cosmological constant problem, and we work under the assumption that a mechanism that sets the contribution from the electroweak minimum of the Higgs potential must be present.} If we had not subtracted the Higgs contribution to the Jordan-frame cosmological constant, the light scalar mass would have received corrections of order $\mu^2$.\footnote{This observation is reminiscent of the ideas behind Higgs-dilaton models \cite{Shaposhnikov:2008xb,Garcia-Bellido:2011kqb}, where the dilaton potential is generated by the Jordan frame cosmological constant in order to realise a quintessence-like scenario.} When the density exceeds the critical density,  the mass of the light mode increases as \smash{$\sim\sqrt{\rho_{\psi}}/M$}. 
It is also important to notice that the strength of the coupling of the light mode to matter perturbations (which will control the strength of the scalar-mediated fifth force) varies with the environmental density and is suppressed when the density exceeds the critical density.

%%%%%%%%%%%%%%%%%%%%

\section{Screening}
\label{sec:screenign}

Considering the equation of motion for the light mode in equation (\ref{eq:scalareom}), we see that the coupling of the light scalar to matter is suppressed in regions of high density (above the critical density $\rho_{\rm crit} =2 \mu_{\chi}^2M^2/(2c-b^2)$). This comes from the variation of $\tilde{\chi}_m$ with the density of the environment.   

The light mode mediates a fifth force, on a test particle with unit mass, of strength
\begin{equation}
    F_{ s} = -\frac{\beta(\rho_{\psi})}{M}\nabla s\;.
    \label{eq:fifth}
\end{equation}
As the coupling $\beta(\rho_{\psi})$ varies with the environment, so will the strength of the scalar-mediated fifth force. This Section explores the phenomenological consequences of this environmental dependence.

%%%%%%%%%%%%%%%%%%%%

\subsection{Environmental screening }

We first consider the situation where the fifth force mediated by the light mode is suppressed because the environment in which we make our observations has a density that exceeds the critical density. 
The characteristic scale over which $s$  can vary is given by the Compton wavelength $\sim 1/m_{\rm eff}$.  
For the fifth force to be suppressed, or screened, the density must exceed the critical density
\begin{equation}
    \frac{\rho_{\psi}}{\mbox{gcm}^{-3}}\gtrsim \frac{0.46}{2c-b^2} \left(\frac{\mu_{\chi}}{\mbox{eV}}\right)^2\left(\frac{M}{\mbox{GeV}}\right)^2\;,
\end{equation}
over a region of spatial extent at least as large as the Compton wavelength.  We should take care when applying this requirement, because in regions of high density (above the critical density), the mass of the light field will increase and the Compton wavelength will shorten. 
Above the critical density,  the coupling function becomes
\begin{equation}
    \beta(\rho_{\psi}) \approx \frac{2b\mu_{\chi}^2M^2}{(2c-b^2)\rho_{\psi}}\;,
\end{equation}
and the coupling is dynamically suppressed compared to our naive expectation of $\beta \sim 1$. 

Two useful examples of this condition for environmental screening are:  
\begin{itemize}
    \item 
The density of the interstellar medium is $\sim 10^{-26}\mbox{ g/cm}^3$.
This exceeds the critical density if 
\begin{equation}
 \frac{1}{(2c-b^2)^{1/2}} \left( \frac{\mu_{\chi}}{\mbox{eV}}\right)  \left(\frac{M}{\mbox{GeV}}\right) \lesssim 2.1 \times 10^{-13}\;.
\label{eq:ismdens}
\end{equation}

\item The density of the Earth is  $\sim 6 \mbox{ g/cm}^3$. This exceeds the critical density  if
\begin{equation}
  \frac{1}{(2c-b^2)^{1/2}}  \left(\frac{\mu_{\chi}}{\mbox{eV}}\right)\left(\frac{M}{\mbox{GeV}}\right) < 5.1 \;.
  \label{eq:earth}
\end{equation}
\end{itemize}
If the first of these conditions is satisfied, we expect the fifth force to be suppressed within the solar system.  If the second, weaker, bound is satisfied, the fifth force will be suppressed within the Earth, but this may not be a sufficient condition to suppress the effects of the scalar in all terrestrial experiments.

%%%%%%%%%%%%%%%%%%%%

\subsection{Thin-shell screening of fifth forces}

If the environment in which an experiment is performed is not dense enough to exceed the critical density, it is still possible that the force sourced by large  objects may be suppressed through the so called `thin-shell' effect \cite{Khoury:2003aq,Khoury:2003rn}.

To see this, we return to working in terms of the field $\tilde{\chi}$, whose background value in a region of density $\rho_{\psi}$ is given by equation (\ref{eq:chimin}). Fluctuations around this value are given by $\delta \chi = s\cos \theta $, and the perturbations $s$ have a density-dependent mass given by equation (\ref{eq:meff}). The thin-shell effect can occur when $\rho_{\rm out} < \rho_{\rm crit}$ but $\rho_{\rm in}> \rho_{\rm crit}$. 

We  consider the profile of the  field around a spherical compact object.  We centre our spherical coordinate system on the centre of the object and  assume that the object has a constant density $\rho_{\rm in}$ when $r\leq R$, where $R$ is the radius of the object.  The object is embedded in a background of constant density $\rho_{\rm out}$. Assuming that $\rho_{\rm in} \gg \rho_{\rm crit}$ and $\rho_{\rm out}\ll \rho_{\rm crit}$, we find
\begin{equation}
    \frac{ \tilde{\chi}}{M}=\left\{\begin{array}{lc}
-\frac{b}{2c-b^2}\left(1-\frac{2(1+m_{\rm out}R)}{m_{\rm in} r}e^{-m_{\rm in}R}\sinh m_{\rm in}r\right)\;, & \mbox{ if } r\leq R\;,\\
-\frac{b\rho_{\rm out}}{2M^2 \mu_{\chi}^2}-\frac{bR}{(2c-b^2)r}e^{m_{\rm out}(R-r)}\;, & \mbox{ if } r>R\;,
    \end{array}\right.
    \label{eq:deltachi}
\end{equation}
where $m_{\rm in}$ and $m_{\rm out}$ indicate the effective mass of the light scalar mode evaluated at the densities $\rho_{\rm in}$ and $\rho_{\rm out}$, respectively. We have additionally assumed that $m_{\rm in}R\gg1$ and will confirm when this condition holds shortly. 

Substituting the expression for the field profile, equation (\ref{eq:deltachi}), around the compact object  into the expression  for the fifth force on a test particle, equation (\ref{eq:fifth}),  we find that the fifth force on a test particle of unit mass at some $r>R$ is 
\begin{equation}
    F_{5}= -\frac{b^2R}{(b^2-2c)r^2}(1+m_{\rm out}r)e^{m_{\rm out}(R-r)}\;.
\end{equation}
This can be compared to the strength of the fifth force of a canonical light scalar with Yukawa couplings to matter controlled by the energy scale $M$ and with mass $m_{\rm out}\approx \mu_{\chi}$, given by
\begin{equation}
    F_{\rm Yuk} = -\frac{M_c}{M^2 r^2}e^{-m_{\rm out}r}\;,
\end{equation}
where $M_c = 4\pi \rho_{\rm in}R^3/3$. We find
\begin{align}
    \frac{F_{ 5}}{F_{\rm Yuk}} & = \frac{b^2RM^2}{(b^2-2c)M_c}(1+m_{\rm out}r)e^{m_{\rm out}R}\;.
    \end{align}
If $m_{\rm out}R\ll 1$ then
\begin{align}
    \frac{F_{ 5}}{F_{\rm Yuk}} & \approx \frac{b^2RM^2}{(b^2-2c)M_c}\nonumber  \\
    & \approx \frac{3b^2(4c-b^2)}{16 \pi(b^2-2c)(m_{\rm in} R)^2}\;,\label{eq:scree}
\end{align}
and the fifth force is suppressed if
\begin{equation}
\frac{M_c}{R}\gg \frac{|b^2-2c|M^2}{b^2}\;.
\label{eq:thinshell}
\end{equation}
When the condition in equation (\ref{eq:thinshell}) is satisfied, the source is sufficiently compact that the fifth force it sources is suppressed and constraints on the model parameters from fifth-force searches are weakened, similarly to chameleon \cite{Khoury:2003aq, Khoury:2003rn} and symmetron models \cite{Hinterbichler:2010es, Hinterbichler:2011ca} (for earlier related work, see Refs.~\citen{Dehnen:1992rr, Gessner:1992flm, Damour:1994zq, Pietroni:2005pv, Olive:2007aj}). Consistency of our solution requires the closely related condition
\begin{equation}
m_{\rm in }^2 R^2=\frac{3(4c-b^2)}{16 \pi M^2}\frac{M_c}{R} \gg1\;.
\end{equation}

As an example, we consider the Earth embedded in the Interstellar Medium (ISM).   We assume there is no environmental screening, so that the density of the ISM does not  exceeds the critical density,  but the density of the Earth does exceed the critical density. This allows for the  possibility of thin-shell screening. Combining equations (\ref{eq:ismdens}) and (\ref{eq:earth}), such a situation can occur when
\begin{equation}
    2.1 \times 10^{-13}\lesssim \frac{1}{(2c-b)^{1/2}}\left(\frac{\mu_{\chi}}{\mbox{eV}}\right)\left(\frac{M}{\mbox{GeV}}\right)\lesssim 5.1\;.
\end{equation}
Then the fifth force sourced by the Earth is screened through a thin-shell effect if equation (\ref{eq:thinshell}) is satisfied for the Earth, which requires
\begin{equation}
    \frac{M}{\mbox{ GeV}}\ll \frac{(3 \times 10^{14})b}{|b^2-2c|^{1/2}} \;.
\end{equation}

%%%%%%%%%%%%%%%%%%%%

\section{Discussion}
\label{sec:dis}

We have presented a theory in which a light scalar field is added to the SM, but the long-range fifth forces mediated by this scalar can be suppressed through environmental screening. The screening occurs because of non-linearities in the scalar potential. In this way, the screening is similar to a number of commonly studied models. The existence of a critical density above which screening can occur means the phenomenology of the theory is particularly similar to that of symmetron models of screening.  However, the key difference between our model and those in the existing literature is the source of the non-linearities.  Prior to this work, it was assumed that screening could only occur if non-linear terms were added to the Lagrangian of the light scalar. Here, we have shown that if the light scalar has a potential which contains only a mass term, and couples to matter through the Higgs portal, then the self-interactions in the Higgs potential are sufficient to induce screening at low energies, where the Higgs field (or, more precisely, the heavy mode of the coupled theory) can be assumed to be non-dynamical. 

In this work, we have only kept terms in the conformal coupling up to order $1/M^2$.  This led to an effective theory, where the mass of the light scalar and its coupling to matter depend on the environmental density. The kinetic term for the light mode in our theory is also re-scaled, as can be seen in equation (\ref{eq:scalareom}), but this is independent of the environment.  It is possible that if we were to extend the calculation to include terms of higher order in $1/M$, we would find environmental dependence occurring in the kinetic sector as well, opening up the possibility of additional Vainshtein-like screening occurring if it becomes more challenging for the light mode to propagate in regions of high density.  

We have  only studied a toy model Lagrangian, with a prototype real-valued Higgs field and a single fermion.  This fails to capture the dynamics of the QCD sector, potentially a significant failing, since the QCD binding energy provides approximately 99\% of the mass of nucleons. In our previous work \cite{Burrage:2018dvt}, following earlier references \cite{Shifman:1978zn, Jungman:1995df, Gunion:1989we},  we showed that an interaction between a conformallly coupled scalar and baryons does arise, mediated through the Higgs field and the conformal anomaly.  This allows baryonic matter to act as a source of energy density in the equations of motion for the light mode of our model in the same way as the fermionic density we have used in the above calculation. Even so, the differing origins of the interactions may lead to an effective violation of the weak equivalence principle between the SM leptons and hadrons~\cite{Burrage:2018dvt}.

It is important to recognise that the analysis presented here considers only tree-level interactions. The potential generated for the light degree of freedom, as described here, will be subject to radiative corrections. A detailed study of these radiative corrections is beyond the scope of this article and may be presented elsewhere. 

In addition, we leave a detailed analysis of experimental constraints for future work.  As well as allowing a theory to avoid existing constraints, screening also introduces novel observational signatures, and these would be smoking-gun signatures of this type of new physics.  For example, long-lived environmentally dependent scalars could have different displaced vertices in the ATLAS and CMS detectors because of the differing design and construction of the detectors \cite{Argyropoulos:2023pmy}.

%%%%%%%%%%%%%%%%%%%%

\section{Conclusions}
\label{sec:conc}

We have studied light conformally coupled scalar fields, a widely considered type of new physics beyond the SM.  We have assumed that the bare potential for these light scalars only contains a mass term. After a series of field redefinitions, we have shown that such a theory is equivalent to a Higgs-portal model, with a particular combination of Higgs-portal couplings.  This combination of couplings may not seem intuitive when viewed as a Higgs-portal model, but we have seen how this arises naturally from the conformal coupling. This, of course, does not preclude large radiative corrections that would require fine-tuning.

In the case of a toy SM, we proceeded by expanding in fluctuations of the scalar fields around a classical minimum of the Einstein-frame potential, and derived the effective equation of motion for the light mode --- the fifth-force mediator of the scalar-tensor theory. Choosing the would-be electroweak minimum to have vanishing potential, so as to eliminate any contribution to the Jordan-frame cosmological constant, the mass of the light mode does not receive large electroweak-scale corrections at tree-level in the Einstein frame.

In all models of screening to date, the screening of fifth forces has occurred because of non-linearities in the equation of motion of the additional scalar field.  In contrast, in the model we study here, there are no terms  in the Lagrangian, which involve  powers of the light field higher than quadratic, and so its  equations of motion are linear.  The only field with non-trivial self interactions is the Higgs.  In the leading semi-classical approximation, we find that these non-linearities are communicated to the fifth-force mediator, resulting in an environmentally dependent effective theory for the light mode.  This environmental dependence appears as density-dependent effective masses and couplings, leading to the screening of the light mode and the fifth force that it mediates.  We find that in different regions of parameter space, the effects of the scalar near the surface of the Earth could be screened by the environment, or by a thin-shell effect. This occurs, despite an absence of non-linear terms in the original Jordan-frame Lagrangian for the light field, as a result of the conformal coupling to the Ricci scalar and the non-linearities in the Higgs potential. 

%%%%%%%%%%%%%%%%%%%%

\section*{Acknowledgments}

This work was supported by a Research Leadership Award from the Leverhulme Trust (CB), a United Kingdom
Research and Innovation (UKRI) Future Leaders Fellowship [Grant No.~MR/V021974/2] (PM), and the Science and Technology Facilities Council (STFC) [Grant Nos. ST/T000732/1 (CB) and ST/X00077X/1 (PM)]. For the purpose of open access, the authors have applied a Creative Commons Attribution (CC BY) licence to any Author Accepted Manuscript version arising.

%%%%%%%%%%%%%%%%%%%%

\section*{Data Access Statement}

No data were created or analysed in this study.

%%%%%%%%%%%%%%%%%%%%

\section*{Author Contribution}
Both authors have contributed equally to the original ideas and calculations in this work, and to the writing of the text. 

%%%%%%%%%%%%%%%%%%%%

\section*{Competing Interests}
Neither author has any competing interest that may affect or be affected by the research reported in this paper. 

%%%%%%%%%%%%%%%%%%%%

\end{document}